\documentclass[prl,aps,showpacs,twocolumn]{revtex4}
\usepackage{graphicx}
\usepackage{amsmath,amssymb,revsymb}
\usepackage[colorlinks=true,citecolor=blue]{hyperref}

\begin{document}

\title{Universal three-body parameter in heteronuclear atomic systems}

\author{Yujun Wang}
\author{Jia Wang}
\author{J. P. D'Incao}
\author{Chris H. Greene}
\affiliation{Department of Physics and JILA, University of Colorado, Boulder, Colorado, 80309-0440, USA}

\begin{abstract}
In Efimov physics, a three-body parameter (3BP), previously regarded as nonuniversal, uniquely defines bound and scattering properties of three particles.  A universal 3BP, however, have been recently shown in experiments and theory in ultracold homonuclear gases. Our present study further predicts a universal 3BP 
for heteronuclear atomic systems near broad Feshbach resonances, and provides physical interpretations for its origin. We show that for a system composed of two light and one heavy atoms, the physical origin of the universal 3BP is similar to the homonuclear case while for systems composed by one light and two heavy 
atoms the universality of the 3BP is now mostly controlled by the heavy-heavy interatomic properties. 
This new universality is explained by the universal properties of the van der Waals interactions in a simple Born-Oppenheimer (BO) picture. 
Finally, we show the numerically determined 3BPs for some combinations of alkali atoms used in ultracold experiments.
\end{abstract}
\maketitle

The Efimov effect~\cite{Efimov}, predicted in the 70's by Vitaly Efimov, has attracted broad interest in studies in atomic and nuclear physics~\cite{Braaten}. 
Motivated by the general interest to understand this esoteric quantum phenomenon and  
its consequences on ultracold quantum gases, significant progress has been made in studies of Efimov physics during the last decade. 
In particular, many theoretical predictions and new
phenomena have been identified due to the extraordinary ability to control the interactions in ultracold quantum gases~\cite{ChinRev}. 

The Efimov effect is characterized by the formation of an infinity of 
trimer states (Efimov states) when the two-body $s$-wave scattering length $a\rightarrow\infty$. The energies of Efimov states, $E_n$, follow 
a universal scaling law: $E_n=E_0 e^{-2n\pi/s_0}$, where $s_0$ is a universal constant  that depends only on the mass ratios, the number of 
resonant pairs of interactions, and the identical particle symmetry. 
The ground Efimov state energy $E_0$, therefore, fully determines the Efimov spectrum and it is usually considered as the three-body parameter (3BP). Other definitions of 3BP exist, however, based on scattering properties of the system. For instance, 
in ultracold quantum gases, the formation of an Efimov state can be manifested by the loss of atoms via three-body recombination via
 resonance ($a<0$) or interference ($a>0$) features in the loss rate $K_3$ as a function of $a$. 
The value of $a$ where the first resonance or interference occurs,  $a_-^*$ or $a_0^*$, respectively, are also equally good 3BPs (relation between $E_0$, $a_-^*$ and $a_0^*$ can be found in Ref.~\cite{Braaten}). Nevertheless, since the early years after Efimov's prediction, 
fundamental assumptions led to the expectation that the 3BP should be nonuniversal \cite{Braaten}.

Despite such expectations,  
ultracold experiments with alkali atoms~\cite{CsExp,KExp,Li7Exp1,Li7Exp2,Li7Exp3,Li6Exp1,Li6Exp2,Li6Exp3,Li6Exp4,RbExp} 
have observed a universal value for $a_-^*$. For homonuclear systems it was found that $a_-^*$$\approx$$-9.1 r_{\rm vdW}$, where $r_{\rm vdW}$ is the van der Waals radius \cite{ChinRev}.
The universal homonuclear 3BP was subsequently discussed by different theoretical studies~\cite{Wang3BP,Chin3BP,Schmidt3BP,Sorensen3BP}. In particular, 
the adiabatic hyperspherical (hereafter ``adiabatic'') picture used by Wang, {\it et al.}~\cite{Wang3BP} shows the existence of an effective three-body repulsion when the hyperradius $R$ 
--- the overall size of a three-body 
system --- is about $2 r_{\rm vdW}$ in homonuclear atomic systems, preventing three atoms from approaching shorter distances. 
The Efimov states are affected by this repulsion, yielding a value for the 3BP that is universally determined 
by $r_{\rm vdW}$~\cite{Wang3BP}. Nevertheless, in 
a heteronuclear system the extension of the universality in 
3BP is not straightforward, particularly for the Efimov states with non-resonant interaction between one of the pairs.

In this Letter, we predict universal 3BP in heteronuclear atomic systems with two identical bosons ($A$) and one distinguishable atom ($X$). 
For a fixed mass ratio, the heteronuclear 3BP is found to depend only on $r_{\rm vdW}$ between the pairs and the homonuclear scattering length. 
We, however, give 
different interpretations for the physical origin of the universality in the ``Efimov-favored'' systems ($s_0>1$) in the extreme of two heavy and one light atoms 
and the ``Efimov-unfavored'' systems ($s_0<1$) in the extreme of two light and one heavy atoms, respectively. 
The universality of the 3BP in the ``Efimov-favored'' systems 
is intuitively understood in the BO approximation, via 
the universal properties of the van der Waals interaction 
between the heavy particles at small distances. 
On the other hand,  
the universality in the ``Efimov-unfavored'' systems will be shown to have similarities with the 
one found in 
homonuclear systems~\cite{Wang3BP}.
Finally, the 3BPs for some experimentally available atomic species are listed in terms of $a_-^*$ and $a_0^*$. 

The wavefunction for the relative motion of three atoms $\Psi$ is determined by the three-body Schr{\"o}dinger 
equation (in a.u.)
\begin{eqnarray}
\left[-\frac{1}{M}\nabla_{\boldsymbol{r}}^2-\frac{2M+m}{4M m}\nabla_{\boldsymbol{\rho}}^2
+V_{AA}(r)+V_{AX}\left(\left|\boldsymbol{\rho}+\frac{\boldsymbol{r}}{2}\right|\right)
\right.\nonumber\\
\left.+V_{AX}\left(\left|\boldsymbol{\rho}-\frac{\boldsymbol{r}}{2}\right|\right)\right]\Psi=E\Psi,
\label{Eq:Schrod}
\end{eqnarray}
where $\boldsymbol{r}$ is the displacement vector between the two $A$ atoms with mass $M$, and 
$\boldsymbol{\rho}$ is the vector from the center of mass of the $A$ atoms to the $X$ atom with mass $m$.
Here 
we use the Lennard-Jones~\cite{LJ} potential that has a van der Waals tail to model the interactions between the atoms with 
distance $r_{AA}$ ($r_{AX}$):
\begin{eqnarray}
V_{AA/AX}(r_{AA/AX})\!=\!-\frac{C_{6,AA/AX}}{r_{AA/AX}^6}\left[1\!-\!\left(\frac{r_{c,AA/AX}}{r_{AA/AX}}\right)^6\right].
\end{eqnarray}
The scattering length 
$a_{AA}$ ($a_{AX}$) is changed by tuning the short-range cut-off $r_{c,AA}$ ($r_{c,AX}$).
We should note, however, that 
our single-channel treatment of the atomic interactions 
 implies that the results presented here are valid for broad Feshbach resonances and less so for narrow ones \cite{NarrowRes}. 

Equation~(\ref{Eq:Schrod}) can be solved to a desired numerical accuracy in the hyperspherical representation~\cite{Suno,WangThesis}, where the 
hyperradius $R$ is defined by
\begin{equation}
\mu R^2=\frac{M}{2} r^2+ \frac{2M m}{2M+m}\rho^2,
\end{equation}
$\mu$ being the three-body reduced mass \cite{RedMass}, and the set of hyperangles $\Omega$ defines the three-body geometry. 
Expanding $\Psi$ in the complete, orthonormal adiabatic basis~\cite{Suno,WangThesis} $\Phi_\nu$ by 
\begin{eqnarray}
\Psi=\sum_\nu F_{\nu,E}(R)\Phi_\nu(R;\Omega),
\end{eqnarray}

Eq.~(\ref{Eq:Schrod}) reduces to a set of coupled hyperradial equations: 
\begin{eqnarray}
&&\!\!\!\left[-\frac{1 }{2\mu}\frac{d^2}{dR^2}+U_{\nu}(R)-E\right]F_{\nu,E}(R)\nonumber\\
&&\!\!\! -\frac{1}{2\mu}\sum_{\nu'} \left[2P_{\nu\nu'}(R)\frac{d}{dR}+Q_{\nu\nu'}(R)\right]F_{\nu',E}(R)=0,
\label{Eq:SchHyper}
\end{eqnarray}
with $P_{\nu\nu'}$ and $Q_{\nu\nu'}$ the nonadiabatic couplings. 
Typically, for well-separated adiabatic potentials $U_\nu(R)$ the spectrum of the three-body system is determined by 
the effective potentials $W_\nu(R)=U_\nu(R)-Q_{\nu\nu}(R)/2\mu$ and the corresponding channel wavefunctions $F_{\nu,E}(R)$. 

For Efimov-favored cases ($M/m\gg 1$) the adiabatic potentials $U_\nu(R)$ are not isolated at small $R$, and the strong nonadiabatic couplings 
near the sharp avoided crossings make an analysis of 3BP based on $W_\nu(R)$ impractical.  These crossings can be removed in the BO approximation where $r$ is taken as the adiabatic variable. 

Here we restrict our discussions only to the $\sigma_g$ symmetry~\cite{Bransden} since the Efimov effect only involves $s$-wave interactions~\cite{Efimov}. 
After a separable form of the wavefunction is assumed, 
\begin{equation}
\Psi=F_{\nu,E}^{\rm BO}(\boldsymbol{r})\Phi_\nu^{\rm BO}(\boldsymbol{r};\boldsymbol{\rho}), 
\end{equation}
the three-body problem reduces approximately to an effective two-body problem, 
and the three-body spectrum can be determined by solving the BO channel wavefunction $F_{\nu,E}^{\rm BO}(r)$ via 
\begin{eqnarray}
\left[-\frac{1}{M}\nabla_{\boldsymbol{r}}^2+U_{\nu}^{\rm BO}(r)-E\right]F_{\nu,E}^{\rm BO}(\boldsymbol{r})=0,
\label{Eq:BO}
\end{eqnarray}
where $U_{\nu}^{\rm BO}(r)$ is the BO potential, obtained by solving Eq.~(\ref{Eq:Schrod}) for fixed values of $r$.
It is well known that when $|a_{AX}|\gg r_{{\rm vdW},AX}$, the potential
relevant to the Efimov effect has the universal long-range behavior~\cite{Braaten}: 
\begin{eqnarray}
U_\nu^{\rm BO}(r)\approx -\frac{\chi_0^2}{2m r^2}\;\; (r_{{\rm vdW},AX} \ll r\ll |a_{AX}|)
\end{eqnarray} 
with $\chi_0\approx 0.567143$. 
In the BO limit, $\chi_0$ is connected to the Efimov scaling constant $s_0$ by $s_0^2\approx \chi_0^2M/2m-1/4$. The 3BP, however, is determined by the behavior of the potential $U_\nu^{\rm BO}(r)$ near the short-range radius $r_0$, 
where $r_0$ is the larger of $r_{{\rm vdW},AX}$ and $r_{{\rm vdW},AA}$. 
The numerical solutions of Eq.~(\ref{Eq:Schrod}) shows that in this region $U_{\nu}^{\rm BO}(r)\approx V_{AA}(r)$ and  since the 3BP is controlled by the position of the last 
node in $F_{\nu,E}^{\rm BO}(r)$ near $r=r_0$~\cite{Braaten}, the property of the potential $V_{AA}(r)$ is the key to 
determine the 3BP. This is in strong contrast to the homonuclear system~\cite{Wang3BP} where the short-range details of the interactions are not referenced at all. 

\begin{figure}
\includegraphics[scale=0.6]{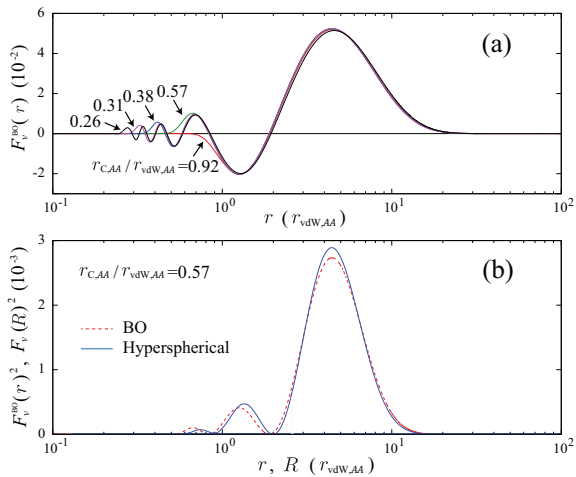}
\caption{
(color online) 
The wavefunctions for the first excited Efimov state at $a_{AA}$=$a_{AX}$=$\infty$.
(a) The BO wavefunction $F_\nu^{\rm BO}(r)$ with increasing number of $A_2$ bound states. The short-range cutoff is tuned at $r_{c,AA}/r_{{\rm vdW}, AA}\approx$0.92, 0.57, 0.38, 0.31, 
0.26 to give 1, 2, 4, 6, 8 $A_2$ $s$-wave bound states, respectively. 
(b) A comparison of the BO probability density $|F_\nu^{\rm BO}(r)|^2$ and hyperspherical probability densities $|F_\nu(R)|^2$ for the first excited Efimov state at unitarity. 
Here we use $A$=$^{174}$Yb and $X$=$^{6}$Li, with $r_{{\rm vdW}, AA}$=78.7 a.u. and $r_{{\rm vdW}, AX}$=38.1 a.u.
 }
\label{Fig:BOWave}
\end{figure}

From the universal van der Waals theory~\cite{Flambaum,Gao} we know that the solution to 
the two-body Schr{\"o}dinger equation with the potential $V_{AA}(r)$ is universally 
determined by $a_{AA}$ and $r_{{\rm vdW},AA}$. The channel wavefunction $F_\nu^{\rm BO}(\boldsymbol{r})$ for the Efimov states 
at short-range $r\lesssim r_0$ therefore also shares 
these universal properties, as demonstrated in Fig.~\ref{Fig:BOWave} for $^{174}$Yb$_2$$^6$Li system with at unitarity ($a_{AX}=a_{AA}=\infty$). 
Here we assume a $J^\Pi=0^+$ symmetry for the total orbital angular momentum $J$ and total parity $\Pi$.
Figure~\ref{Fig:BOWave}(a) shows that for fixed $a_{AA}$, 
decreasing the short-range cut-off $r_{c,AA}$ (consequently creating more $A_2$ bound states) only builds more oscillations in $F_\nu^{\rm BO}(r)$ at small $r$ without changing its long-range behavior. In fact,
the universality of the 3BP is supported by the position of the last node in $ F_{\nu,E}^{\rm BO}(r)$ (and therefore the corresponding energy), 
which changes only within a few percent when the depth of the potential 
$V_{AA}(r)$ is changed by more than 3 orders of magnitude 
for values of $r_{c,AA}$ shown in Fig.\ref{Fig:BOWave}(a). 

Although the BO approximation should yield the exact 3BP in the limit $M/m\rightarrow\infty$, it is important to know the significance of the 
non-BO correction in realistic atomic systems. The role of the $A_2+X$ break-up channels in determining the 3BP is of particular interest due to 
the inability of the BO representation to describe such channels. To this end, we solve Eq.~(\ref{Eq:SchHyper}) to obtain the exact 
Efimov spectrum. 
Figure~\ref{Fig:BOWave}(b) shows a comparison of the 
BO and the hyperspherical radial probability densities ($|F_\nu^{\rm BO}(r)|^2$ and $|F_\nu(R)|^2$) for the first excited Efimov state at unitarity. The good agreement shown in Fig.~\ref{Fig:BOWave}(b) was observed numerically for mass ratio $M/m$ ranging from 29 to 14, corresponding to Yb$_2$Li and Rb$_2$Li systems ~\cite{Unpub}.

\begin{figure}
\includegraphics[scale=0.55]{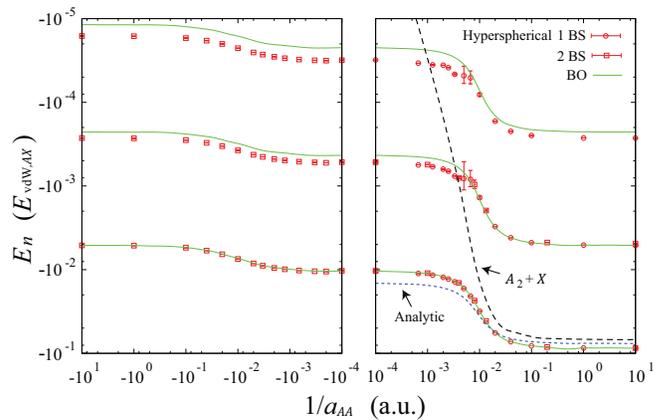}
\caption{
(color online) 
The Efimov energy spectrum $E_n(a_{AA})$ for $^{174}$Yb$_2$$^6$Li at the heteronuclear unitarity ($a_{AX}$=$\infty$). The ``1BS'' and ``2BS'' hyperspherical calculations 
have different $r_{c,AA}$ and $r_{c,AX}$ to give 1 and 2 $s$-wave bound states for both $A+A$ and $A+X$ pairs, respectively. The error bar on the hyperspherical data points indicates 
the width of the corresponding Efimov state. The ``Analytic'' curve is calculated from Eq.~(\ref{Eq:E0}) (see the discussion in the main text). 
 }
\label{Fig:EfimovSpec}
\end{figure}

To study the non-BO correction to the Efimov energy spectrum more quantitatively, we first recall~\cite{Flambaum,Gao} that 
the position of the last node in $F_\nu^{\rm BO}(r)$ for a pure van der Waals interaction also depends on $a_{AA}$, suggesting a universal 
dependence of the 3BP on $a_{AA}$. Figure~\ref{Fig:EfimovSpec} shows such dependence for 
the low-lying Efimov states for the $^{174}$Yb$_2$$^6$Li system. 
The spectrum is calculated by both the BO approximation and the hyperspherical representation. The energy of the ground Efimov state calculated by the BO approximation agrees almost perfectly with the exact hyperspherical calculations. (The ground Efimov state is defined by 
the first Efimov state that appears at $a_{AX}=a_-^*$.) 
The energies for the 
excited states, however, start to deviate due to the finite mass ratio correction to the Efimov scaling factor $s_0$ in the BO calculations. 
Nevertheless, the overall agreement still
provides further evidence for universality of the 3BP. 

The Efimov spectrum shown in Fig.~\ref{Fig:EfimovSpec} is periodic 
in $1/a_{AA}$, 
which implies that as the depth of the two-body potentials $V_{AA}$ 
increases so that $a_{AA}$ changes periodically from negative to positive, an Efimov state will follow the path in the spectrum and gets more deeply bound 
until it moves below the heteronuclear van der Waals energy scale and becomes non-Efimov.

The role of the $A_2+X$ break-up channel is shown in Fig.~\ref{Fig:EfimovSpec} for the cases where $a_{AA}\gg r_{0}$. 
Evidently, all the excited Efimov states acquire large widths when the first excited state moves right above the $A_2+X$ threshold. In fact, as will be discussed later, in general this threshold energy 
is a dividing point between two different regions of the Efimov energy spectrum: a lower Efimov spectrum controlled by larger $s_0^*$ (corresponding to three resonant interactions) and 
a upper Efimov spectrum controlled by smaller $s_0$ (corresponding to two resonant interactions)~\cite{DIncao,Rittenhouse}. But this division goes away in the BO limit because 
the values of $s_0$ and $s_0^*$ becomes identical~\cite{DIncao}. 

The simple scaling behavior of the BO potential $U_\nu^{\rm BO}(r)$ allows us to determine the 3BP for the Efimov-favored systems analytically. 
The key step to determine the 3BP, in our case here $E_0$, 
is to find the position of the first node in $F_\nu^{\rm BO}(r)$ in the Efimov region $r > r_0$. 
To this end, we divide the BO potentials in the different regions and use their approximate forms. The first region is the van der Waals region with $r<r_-$ where 
$V_{AA}(r)$ is dominant, and we let $U_\nu^{\rm BO}(r)=-C_{6,AA}/r^6$. The second region $r_-<r<r_+$ characterizes the deviation of the potential from both its short-range and 
asymptotic forms, and is approximated as $U_\nu^{\rm BO}(r)=-C_{6,AA}/r^6-(\alpha\chi_0)^2/2mr^2$, where $\alpha^2\approx 2$ from numerical observations. 
In the last region we let $U_\nu^{\rm BO}(r)=-\chi_0^2/2mr^2$. The zero-energy solution in each region can be written down analytically 
in terms of Bessel's functions $J_\nu(x)$ and $N_\nu(x)$, and will be matched at the boundaries. For simplicity, we let $r_-$ and $r_+$ be nodal positions of these solutions, 
leading to the following transcendental equations: 
\begin{eqnarray}
 \frac{J_{-\frac{i\alpha s_0}{2}}\left(2\frac{r_{{\rm vdW},AA}^2}{r_+^2}\right)}{J_{-\frac{i\alpha s_0}{2}}\left(2\frac{r_{{\rm vdW},AA}^2}{r_-^2}\right)}
=\frac{N_{-\frac{i\alpha s_0}{2}}\left(2\frac{r_{{\rm vdW},AA}^2}{r_+^2}\right)}{N_{-\frac{i\alpha s_0}{2}}\left(2\frac{r_{{\rm vdW},AA}^2}{r_-^2}\right)}. 
\end{eqnarray}
\begin{eqnarray}
\frac{N_{\frac{1}{4}}\left(2\frac{r_{{\rm vdW},AA}^2}{r_-^2}\right)}{J_{\frac{1}{4}}\left(2\frac{r_{{\rm vdW},AA}^2}{r_-^2}\right)}
=1-\sqrt{2}\frac{a_{AA}}{r_{{\rm vdW},AA}}\frac{\Gamma(5/4)}{\Gamma(3/4)}, 
\end{eqnarray}
and the expression for $E_0$: 
\begin{equation}
E_0=-\frac{4}{M r_+^2}\exp{\left( -\frac{2}{s_0}\{{\rm Arg}[\Gamma(1-i s_0)]-\pi\} \right )}.
\label{Eq:E0}
\end{equation}
As shown in Fig.~\ref{Fig:EfimovSpec}, Eq.~(\ref{Eq:E0}) gives a reasonable estimate of $E_0$ when compared to the exact numerical results. 

For the Efimov-unfavored cases ($M/m\lesssim 1$) the BO picture becomes invalid. 
In contrast to the Efimov-favored cases, our hyperspherical calculations now show strong nonadiabatic corrections, 
suggesting a different physical regime for the universality in the 3BP. 
 The energy spectrum is now well separated into two parts by the $A_2+X$ threshold energy 
when $|a_{AX}|\gg a_{AA}\gg r_0$. The upper and lower parts of the spectrum follow the Efimov geometric scaling for two- ($s_0$) or three-resonant-interaction ($s_{0}^*>s_{0}$), respectively~\cite{Rittenhouse,DIncao}. 
Therefore in contrast to the Efimov-favored cases there are two 3BP to be considered. 
The two distinct 3BP are manifested by the two well-separated Efimov potentials $W_\nu(R)$ which become more weakly coupled  
as $M/m$ decreases. 
Specifically, the upper Efimov spectrum comes from the three-body potential associated to the $AX+A$ channel, whereas the lower spectrum originates from the potential associated to the $A_2+X$ channel. 

\begin{figure}
\includegraphics[scale=0.55]{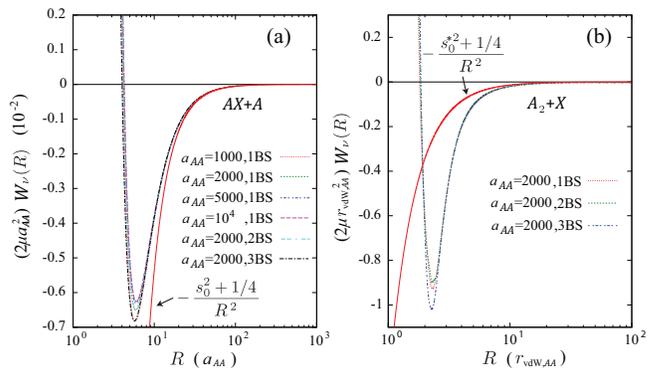}
\caption{
(color online) 
The adiabatic hyperspherical potentials $W_{\nu}(R)$ for the $AX+A$ channel (a) and the 
$A_2+X$ channel (b) for an Efimov-unfavored system with $A$=$^{133}$Cs, $X$=$^{87}$Rb. Here $r_{{\rm vdW}, AA}$=101 a.u. 
and $r_{{\rm vdW}, AX}$=90.9 a.u.. The $a_{AA}$ labels in the plots are also in a.u..
 }
\label{Fig:UnfavorPot}
\end{figure}

To simplify our discussion, the analysis below is made only for 
$a_{AX}=\infty$, however, same conclusions hold for finite values of $a_{AX}$. Figure~\ref{Fig:UnfavorPot}(a) shows the universal behavior of the $AX+A$ 
Efimov potential $W_\nu(R)$ for $^{133}$Cs$_2$$^{87}$Rb. The universal form of the potentials after a scaling by $a_{AA}$ strongly suggests a universal 3BP 
that depends only on $a_{AA}$. 
Figure~\ref{Fig:UnfavorPot}(b) shows the universal properties of the $A_2+X$ Efimov potential, which resemble those in the homonuclear case~\cite{Wang3BP}. In this case, $W_\nu(R)$ also displays a repulsive barrier near $R\gtrsim r_0$ independent of the number of two-body bound states .
It is also important to note that when $a_{AA}\ll r_0$ there is only one continuous Efimov spectrum 
with scaling constant $s_0$ and the Efimov potential $W_\nu(R)$ shares the same universal features as the potentials shown in Fig.~\ref{Fig:UnfavorPot}(b).
Although the Efimov potentials are universal for the Efimov-unfavored cases, we should point out that the exact shape of the potentials near their minimum depends on the mass ratio $M/m$.

Table~\ref{Tab:3BP} summarizes our results obtained for the 3BPs relevant to the three-body recombination process, $A+A+X\rightarrow AX+A$~\cite{Petrov,DIncao}, for some combinations of alkali atoms. The 3BPs are given 
in the form of the experimental observables $a_-^*$ and $a_0^*$ in $a_{AX}$  near a $A+X$ Feshbach resonance. 
It is important to note that the knowledge of these universal 3BPs permits more precise calibration of the 
position of a Feshbach resonance in ultracold quantum gases, providing great benefits to more accurate measurements on strongly interacting 
quantum gases. 
Finally, we have also tested the universality of the 3BPs against three-body forces ~\cite{Soldan, Marinescu} and have found a negligible effect on the 3BPs. 
\begin{table}
\begin{ruledtabular}
\begin{tabular}{lccccc}
 & $s_0$ & $s_0^*$ & $a_{AA,bg}$ (a.u.) & $a_0^*$ (a.u.)& $a_-^*$ (a.u.) \\
\hline
$^{174}$Yb$_2$$^6$Li & 2.246 & 2.382 & 104~\cite{YbScatt,YbScatt2} & $1.3\times 10^3$ & $-8.4\times 10^2$\\
$^{133}$Cs$_2$$^6$Li & 1.983 & 2.155 & 2000~\cite{CsScatt} & $9.6\times 10^2$ & $-1.4\times 10^3$ \\
$^{87}$Rb$_2$$^6$Li & 1.633 & 1.860 & 100~\cite{RbScatt} & $3.8\times 10^2$ & $-1.6\times 10^3$  \\
$^{41}$K$_2$$^6$Li & 1.154 & 1.477 & 62~\cite{KScatt} & $3.7\times 10^2$ & $-2.4\times 10^3$ \\
$^{23}$Na$_2$$^6$Li & 0.875 & 1.269 & 100~\cite{NaScatt} & $1.5\times 10^3$ & $-1.3\times 10^4$ \\
$^{87}$Rb$_2$$^{40}$K & 0.653 & 1.125 & 100 & $2.8\times 10^3$ & $<-3\times 10^4$\\
$^{133}$Cs$_2$$^{87}$Rb & 0.535 & 1.060 & 2000 & $2.3\times 10^3$ & $<-4\times 10^4$\\
$^{41}$K$_2$$^{87}$Rb & 0.246 & 0.961 & 62 & $>7\times 10^3$ & $<-1\times 10^6$
\end{tabular}
\end{ruledtabular}
\caption{
The universal Efimov scaling constants $s_0$, $s_0^*$ and the 3BPs $a_{AX}=a_0^*$ and $a_{AX}=a_-^*$ obtained by keeping $a_{AA}$ fixed at its background value ($a_{AA,bg}$).
 } 
\label{Tab:3BP}
\end{table}
It has been predicted~\cite{Petrov} that $a_-^*/a_0^*=e^{\pi/2s_0}$ in the zero-range limit with 
$a_{AA}=0$. 
However, due to the complicated interplay of those finite length scales in realistic systems~\cite{Unpub}, 
significant deviations of this relation 
are observed in the numerical results in Table~\ref{Tab:3BP}. 

It is also worth mentioning that our results for Rb$_2$K and estimations on 3BP for K$_2$Rb 
do not agree with the experimental results by G. Barontini {\it et al.}~\cite{Barontini}.
We believe such disagreement might be related to effective-range corrections and requires 
further investigations. 
Our Rb$_2$K results, however, are more consistent with JILA experiment~\cite{JILAExp}, where no Efimov resonances were observed for $a_{AX}>-10^4$, although a final conclusion cannot be made due to the thermal 
saturation in that experiment.

To summarize, we have predicted and analyzed universality in the 3BP in heteronuclear atomic systems. The physical origin of this universality, however, is different depending on the mass ratio between the atoms. For the Efimov-favored cases, the universality is explained 
by the universal property of the atomic van der Waals interaction between the heavy atoms, which 
yields an analytical determination of the 3BP in the BO limit. For the 
Efimov-unfavored cases, the universality is manifested by a universal 
effective three-body repulsion in the range of the van der Waals radius 
shielding three atoms from the complicated interactions at shorter range.  
Finally, the 3BPs we have calculated can be used in ultracold experiments for calibration of the positions of the Feshbach resonances.

\begin{acknowledgments}
The authors acknowledge stimulating discussions with P. S. Julienne, and
J. M. Hutson. This work is supported in part by the U.S. National Science Foundation and in part by the AFOSR-MURI.
\end{acknowledgments}


\begin{thebibliography}{33}

\bibitem{Efimov} V. Efimov, Sov. J. Nucl. Phys. {\bf 12}, 589 (1971); {\bf 29}, 546 (1979); Nucl. Phys. {\bf A210}, 157 (1973).

\bibitem{Braaten} E. Braaten and H.-W. Hammer, Phys. Rep. {\bf 428}, 529 (2006). 

\bibitem{ChinRev} C. Chin, R. Grimm, P. Julienne, and E. Tiesinga Rev. Mod. Phys. {\bf 82}, 1225 (2010).

\bibitem{CsExp} M. Berninger, A. Zenesini, B. Huang, W. Harm, H.-C. N{\"a}gerl, F. Ferlaino, R. Grimm, P. S. Julienne, and J. M. Hutson, Phys. Rev. Lett. {\bf 107}, 120401 (2011)

\bibitem{KExp} M. Zaccanti, B. Deissler, C. D'Errico, M. Fattori, M. Jona-Lasinio, S. M{\"u}ller, G. Roati, M. Inguscio, and G. Modugno, Nature Phys. {\bf 5}, 586 (2009).

\bibitem{Li7Exp1} S. E. Pollack, D. Dries and R. G. Hulet, Science {\bf 326}, 1683 (2009).

\bibitem{Li7Exp2} N. Gross, Z. Shotan, S. Kokkelmans, and L. Khaykovich,
Phys. Rev. Lett. {\bf 103}, 163202 (2009).

\bibitem{Li7Exp3} N. Gross, Z. Shotan, S. Kokkelmans, and L. Khaykovich,
Phys. Rev. Lett. {\bf 105}, 103203 (2010).

\bibitem{Li6Exp1} T. B. Ottenstein, T. Lompe, M. Kohnen, A. N. Wenz,
and S. Jochim, Phys. Rev. Lett. {\bf 101}, 203202 (2008). 

\bibitem{Li6Exp2} T. Lompe, T. B. Ottenstein, F. Serwane, K. Viering, A. N. Wenz, G. Z{\"u}rn, and S. Jochim, Phys. Rev. Lett. {\bf 105}, 103201 (2010).

\bibitem{Li6Exp3} J. H. Huckans, J. R.Williams, E. L. Hazlett, R.W. Stites
and K. M. OHara, Phys. Rev. Lett. {\bf 102}, 165302 (2009).

\bibitem{Li6Exp4} J. R. Williams, E. L. Hazlett, J. H. Huckans, R. W. Stites, Y. Zhang, and K. M. O’Hara, Phys. Rev. Lett. {\bf 103}, 130404 (2009).

\bibitem{RbExp} R. J. Wild, P. Makotyn, J. M. Pino, E. A. Cornell, and
D. S. Jin, Phys. Rev. Lett. {\bf 108}, 145305 (2012).

\bibitem{Wang3BP} J. Wang, J. P. D'Incao, B. D. Esry, and C. H. Greene, Phys. Rev. Lett. {\bf 108}, 263001 (2012). 

\bibitem{Chin3BP} C. Chin, arXiv:1111.1484 (2011). 

\bibitem{Schmidt3BP} R. Schmidt, S. P. Rath, and W. Zwerger, arXiv:1201.4310 (2012). 

\bibitem{Sorensen3BP} P. K. S{\o}rensen, D. V. Fedorov, A. S. Jensen, and N. T. Zinner, arXiv:1206.2274 (2012). 

\bibitem{LJ} L. E. Jones, Proc. R. Soc. Lond. A {\bf 106}, 463 (1924).

\bibitem{NarrowRes} D. S. Petrov, Phys. Rev. Lett. {\bf 93}, 143201 (2004);
Y. Wang, J. P. D'Incao, and B. D. Esry, Phys. Rev. A {\bf 83}, 042710
(2011).

\bibitem{Suno} H. Suno, B. D. Esry, C. H. Greene, and J. P. Burke, Jr., Phys. Rev. A {\bf 65}, 042725 (2002). 

\bibitem{WangThesis} Y. Wang, Universal Efimov Physics in Three- and
Four-body Collisions, PhD Thesis, Kansas State University, 2010.

\bibitem{RedMass}  For a three-body system with two identical particles ($A$) with mass $M$ and a distinguishable particle ($X$) with mass $m$ 
the three-body reduced mass $\mu=[M^3/2(2M+m)]^{1/2}$ is 
defined here such that it reduces to the reduced mass between the $A$ atoms in the limit $M\gg m$.

\bibitem{Bransden} B. H. Bransden and C. J. Joachain, {\it Physics of Atoms and Molecules}, 2nd Ed., (Prentice Hall, Essex, England, 2003), page 487.

\bibitem{Flambaum} G. F. Gribakin and V. V. Flambaum, Phys. Rev. A {\bf 48}, 546 (1993).

\bibitem{Gao} B. Gao, Phys. Rev. A {\bf 58}, 1728 (1998).

\bibitem{Unpub} Y. Wang, J. Wang, J. P. D'Incao and C. H. Greene, in preparation.

\bibitem{DIncao} J. P. D'Incao and B. D. Esry, Phys. Rev. Lett. {\bf 103}, 083202 (2009).

\bibitem{Petrov} K. Helfrich, H.-W. Hammer, and D. S. Petrov, Phys. Rev. A {\bf 81}, 042715 (2010). 

\bibitem{Rittenhouse} S. T. Rittenhouse, N. P. Mehta, and C. H. Greene, Phys. Rev. A {\bf 82}, 022706 (2010). 

\bibitem{Soldan} P. Sold\'an, M. T. Cvita\v{s}, and J. M. Hutson, Phys. Rev. A {\bf 67}, 054702 (2003). 

\bibitem{Marinescu} M. Marinescu, and A. F. Starace, Phys. Rev. A {\bf 55}, 2067 (1997).

\bibitem{YbScatt} M. Kitagawa, K. Enomoto, K. Kasa, Y. Takahashi, R. Ciury{\l}o, P. Naidon, and P. S. Julienne, Phys. Rev. A {\bf 77}, 012719 (2008).

\bibitem{YbScatt2} K. Enomoto, M. Kitagawa, K. Kasa, S. Tojo and Y. Takahashi, Phys. Rev. Lett. {\bf 98}, 203201 (2007). 

\bibitem{CsScatt} C. Chin, V. Vuleti\'c, A. J. Kerman, S. Chu, E. Tiesinga, P. J. Leo, and C. J. Williams, Phys. Rev. A {\bf 70}, 032701 (2004). 

\bibitem{RbScatt} J. M. Vogels, C. C. Tsai, R. S. Freeland, S. J. J. M. F. Kokkelmans, B. J. Verhaar, and D. J. Heinzen, Phys. Rev. A {\bf 56}, R1067 (1997).

\bibitem{KScatt} H. M. J. M. Boesten, J. M. Vogels, J. G. C. Tempelaars, and B. J. Verhaar, Phys. Rev. A {\bf 54}, R3726 (1996).

\bibitem{NaScatt} A. J. Moerdijk, B. J. Verhaar, and A. Axelsson, Phys. Rev. A {\bf 51}, 4852 (1995). 

\bibitem{Barontini} G. Barontini, C. Weber, F. Rabatti, J. Catani, G. Thalhammer, M. Inguscio, and F. Minardi, Phys. Rev. Lett. {\bf 103}, 043201 (2009).

\bibitem{JILAExp} T. D. Cumby, R. Shewmon, Ming-Guang Hu, and D. S. Jin, private communication. 

\end{thebibliography}
\end{document}